# Longitudinal Monitoring of Periodontal Inflammation by Quantitative Ultrasound: Attenuation and Backscatter Intensity Signature

Daria Poul, *PhD*, Amanda Rodriguez Betancourt, *MS, DDS*, Ankita Samal, *BDS*, Carole Quesada, *BS*, Ted Lynch, *PhD*, Cristel Baiu, *MS*, J. Brian Fowlkes, *PhD*, Hsun-Liang Chan, *MS, DDS*, Oliver D. Kripfgans, *PhD*

**Abstract**. In the United States, periodontal diseases, a spectrum of inflammatory conditions, affect 4 out of 10 adults ($\geq 30$ years). Current standards of care in clinical assessment of these diseases are invasive, subjective, only semi-quantitative, and primarily detecting later stages. Applications of ultrasonography in periodontology has been emerging in recent years. Despite such growing interest, Quantitative ultrasound (QUS) approaches remain largely unexplored and their utility for longitudinal characterization of oral inflammation has yet to be established. Here, we present one of the early investigations into the potential of QUS techniques for inflammation monitoring. In a staggered study involving a pig cohort (N = 8), interdental gingival tissues at three interproximal oral sites (Premolar 3 - Distal, Premolar 4 - Distal, and Molar 1 - Distal) from four quadrants were enrolled. The study involved intraoral ultrasonography at baseline and five inflammation timepoints. Inflammation was induced using two complementary approaches at each site. Two QUS parameters of ultrasound attenuation coefficient slope (ACS) and backscatter intensity (BSI) were investigated. Sex and oral sites were also used to stratify the longitudinal QUS estimates with inflammation. Our results showed that both ACS and BSI were statistically significant from the healthy baseline ($p \leq 0.05$) across all oral sites at week 2 and/or 4 of considering combined sexes. Overall, inflammation inoculation was associated with a decrease in ACS and an increase in BSI within the same investigated regions of interest. Sex-based stratifications of BSI revealed that week4 and/or week 2 remained statistically significant across male and female cohorts at all three oral sites. ACS variations were spread across oral sites and sexes, with Premolar 4 - Distal as overall non-significant. The 2D classification accuracy for distinguishing baseline from week 2, were 92.0%, 82.0%, and 74.2% for M1-Dis, PM4-Dis, and PM3-Dis, respectively This study is among early implementation of QUS approaches for periodontal inflammation characterization. The findings highlight the promising potential of intraoral ultrasonography combined with QUS-based techniques for non-invasive assessment and quantification of oral inflammation. QUS-derived parameters may serve as non-invasive and quantitative biomarkers, offering new approaches for improving the diagnosis in periodontal and dental healthcare.

## I. Introduction

Periodontal diseases are a spectrum of inflammatory conditions affecting soft tissues, i.e. gum, that support teeth such as gingiva, alveolar lining mucosa and alveolar bone, i.e. jawbone. Periodontal diseases span an early and reversible inflammatory phase termed *gingivitis* to advanced stages of *periodontitis* which is irreversible. Periodontitis affects four out of 10 adults aged 30 years or higher in the United States [1]. Figure 1 presents an illustration of oral soft and hard tissues, comparing healthy versus periodontitis conditions. As shown in this figure, periodontitis directly affects oral soft tissues and is characterized with gum tissue recession and attachment loss from tooth and pocketing. Its further progression may advance to bone loss and eventually, may lead to tooth loss. Periodontitis is reported to be associated with systemic effects such as diabetes and cardiovascular diseases [2]. Gingivitis, on the other hand, is characterized by the gingival inflammation with no clinical sign of attachment loss and radiographic bone loss. Thus, an accurate diagnosis of periodontal diseases, particularly at early inflammation stage, can prevent irreversible outcomes which results in enhancing public health and alleviating economic burdens from any necessary and advanced clinical intervention procedures (such as regenerative surgeries, root canals or dental implants that otherwise are imposed on population and healthcare system.

### A. Periodontal Inflammation Assessment in Clinics

Currently, standard of care in clinic for periodontal inflammation monitoring is bleeding on probing (BOP) [3]. In BOP, inflammation is assessed by inserting a marked

Corres. Author 1: D.P.; (ssheykho@umich.edu); Corres. Author 2: O.D.K. (greentom@umich.edu); Corres. Author 3: H-L.C.(chan.1069@osu.edu).

*D.P.*, *C.Q.*, *O.D.K.,* and *J.B.F.*: Department of Radiology, University of Michigan, Ann Arbor, MI, USA. *A.R.B.*: Department of Periodontics, College of Dentistry, University of Illinois Chicago, Chicago, IL, USA. *A.S.*: Department of Periodontics, University of Iowa College of Dentistry & Dental Clinics, Iowa City, IA, USA. *T.L.*: Sun Nuclear Corporation, Norfolk, VA, USA. *C.B.*: Department of Medical Physics, University of Wisconsin-Madison, Madison, WI, USA. *O.D.K.,* and *J.B.F.*: Department of Biomedical Engineering, University of Michigan, Ann Arbor, MI, USA. *H-L.C.*: Division of Periodontology, College of Dentistry, The Ohio State University, Columbus, OH, USA.

periodontal probe into the space between gum and tooth (i.e. gingival sulcus (healthy condition) or periodontal pocket (disease condition), annotated in Figure 1, and reading the penetration depth and observed amount of periodontal bleeding. Clinically, gingiva is considered healthy in the absence of bleeding as well as not observing any edema (swelling), erythema (redness), suppuration, and by far, lack of any recession (attachment loss) or bone loss [4]. BOP relies on the assumption that the degree of periodontal inflammation is positively correlated with penetration depth and potential bleeding due to disruption of microvasculature. Although an absence of BOP strongly indicates absence of inflammation, the presence of BOP is not necessarily a strong indication of disease activity. One underlying reason is that BOP is subjective and operator-dependent in which the penetration depth relies on the size of probe tip, probe insertion angle, and penetration force exerted, affecting sensitivity and reproductivity of measurements. Particularly, BOP may result in false-positive readings associated with excessive probing force greater than 0.25 Newton [5]. BOP is also invasive, unpleasant for patients, only semi-quantitative and a late indicator of inflammation stages [3]. A visual examination is another diagnostic method for inflammation assessment in which swollen and erythematous gum tissues are normally considered as indicative of inflammation. However, it is only qualitative and subjective as thick or pigmented gums can easily hide clinical inflammation signs. Radiographic examinations such as X-ray, cone-beam computed tomography (CBCT) are other clinical approach that complements oral diagnosis. These are mainly employed in evaluation of bone structures such as bone level (or bone loss), as well as assessing the presence of lesions. Notably, X-ray does not provide high-resolution images of oral soft tissues such as gingiva and epithelium. Its accuracy is also compromised due to possible superimposition of oral structures projected onto the 2D imaging plane.

### *B. Ultrasonography for Periodontal/Dental Research*

Given the limitations of current standard of care, there is a clinical gap for a non-invasive, subjective and quantitative diagnostic method for oral inflammation monitoring. To address this gap as well as advancing diagnostic practices in dentistry, research on extending the application of ultrasonography to dentistry has been growing within the past decade [6]. Ultrasound imaging is a non-invasive, non-ionizing, portable and relatively inexpensive modality with its conventional gray-scale B-mode that images tissues' landmark structures [7, 8]. It also offers several quantitative imaging modes in addition to its signature B-modes which have shown promising potentials in tissue characterization. These include but are not limited to quantitative ultrasound imaging for tissue microstructure characterization [9-13],ultrasound-based elasticity imaging [14-17] and ultrasound blood flow imaging [18] established in various biological soft tissues such as liver [19-21], breast [22-24], kidney [25, 26], placenta [27, 28], and muscles [29]. However, their applications in periodontal tissue characterization and disease monitoring are largely unexplored and are still in early stages (summarized below). This delay partly stems from the limitations in adopting available commercial ultrasound probes for periodontal tissue imaging [30, 31].

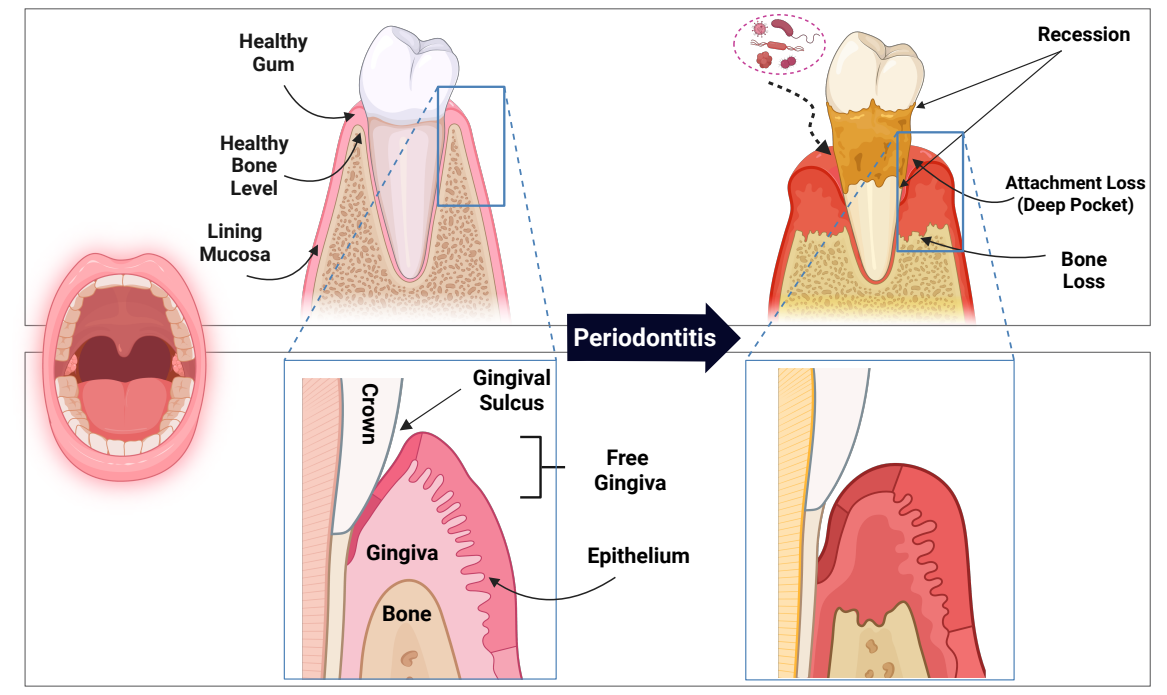

**Figure 1.** Illustration of oral tissues' anatomy at healthy versus periodontitis.

**Clinical Measurements**: Ultrasonography has been used for anatomical measurements of oral soft and hard tissues that have clinical meaning such as thickness/height of gingiva [31-34], mucosa [35], epithelium [36] and rete pegs [37].

**Blood Flow Analysis**: Ultrasound-based blood flow quantification has also been employed in periodontal research. Tavelli et al. [38] employed color flow in a clinical study to monitor wound healing at implant sites. Samal et al. [39] investigated periodontal inflammation using color flow velocity and power in pigs. Barootchi et al. [40] showed correlation between clinically-assessed inflammation with color velocity and power at healthy versus diseased implants.

**Stiffness Analysis**: Yu et al. [41] compared biomechanical properties of gingiva versus alveolar mucosa using tensile strain analysis in a porcine cohort. Tavelli et al. characterized tissue stiffness at implant sites using strain-based elasticity imaging [42]. Woo et al. compared minimum gingival compressional strains required to halt the measured blood flow at baseline versus inflammation to track changes [43].

***B-mode Image Processing***: A group of studies have focused on evaluating image pixel analysis of compressed B-modes such as comparing relative echogenicity from echo levels (decibels), i.e. image analysis. For example, in 2026 Estrade et al. [44], in a clinical study to analyze periodontitis, characterized healthy versus periodontitis by pixel intensity analysis of compressed B-mode images. Galarraga-Vinueza et al.[45] characterized healthy vs. peri-implantitis tissues based on image echo intensity and the first and second order statistical grey level image analysis. Di Stasio et al. [46] used image-based measurements of echo levels (in decibels) directly obtained from compressed B-modes to characterize different oral tissue types. While image-based analyses are characteristic, informative, and quantitative, it is critical to highlight that some information on tissue microstructures may be masked in these images from advanced processing applied by scanners' propriety algorithms to enhance image representation. Moreover, the specific function used by scanners for dynamic range compression of echo envelopes could bias an image-based (gray-value) analysis. **Quantitative Ultrasound (QUS)***:* QUS techniques rely on investigating absolute uncompressed echo signals coming from tissue microstructures and their applications in periodontal tissue characterization are almost entirely unexplored other than a few recent studies. For example, we have recently reported the first implementation of QUS analysis based on speckle statistics modelling for periodontal tissue characterization [47, 48]. We have also reported the first standard quantification of ultrasound attenuation, a key acoustic property of oral soft tissues and a characteristic QUS parameters [49]. It is noted that a 2022 study by Di Stasio et al. [46] reported a measurement of a parameter attributed to attenuation, however, what was actually measured was compressed B-mode image analysis of oral tissues rather than an uncompressed signal analysis. Such attenuation estimation approach deviates from standardized techniques of physical acoustics. Here, for the first time to our knowledge, we examined periodontal tissue changes in response to inflammation using two significant QUS parameters in a longitudinal in vivo preclinical model. These parameters are ultrasound attenuation coefficient slope (ACS) and backscatter intensity (BSI). Attenuation has long been studied for liver characterization [50, 51] and currently is employed in commercialized scanners for liver fat quantification in clinical settings [52]. It has also been investigated in placenta [27], breast [53, 54], pancreas [55], and bone [56]. QUS may offer potential biomarkers for quantitative assessment of periodontal inflammation to complement current standard of care in dentistry.

## II. Materials and Methods

### A. QUS Methods

To estimate the attenuation coefficient, we employed the spectral difference method (SDM), a well-established technique that employs a reference tissue-mimicking phantom to eliminate biases associated with transducer diffraction effects [57, 58]. In a pulse-echo ultrasound imaging system with the transducer transmit pulse of $P(f)$, the echo $S(f,z)$ received at the transducer from a ROI at the depth of $z$ can be modeled as in equation (1) ($f$: frequency).

$$S_s(f,z) = P(f) \cdot B(f,z) \cdot D(f,z) \cdot A(f,z) \qquad (1)$$

$A(f,z)$ is the cumulative attenuation, $D(f,z)$ is the diffraction effect, $B(f,z)$ shows the backscatterinf effect, and $P(f)$ represents the combined effect of the transducer receive sensitivity and transmit pulse power. $A(f,z)$ is modelled as an exponential function $A(f,z) = A(f,z_0)e^{-4\alpha(f)(z-z_0)}$. $A(f,z_0)$ represents the cumulative attenuation coefficient from the imaging start depth to the start depth of the ROI ($z_0$). $\alpha(f)$ is the local attenuation coefficient within the ROI and is a function of frequency, commonly modeled linearly as $\alpha(f) = \beta \cdot f$, with a slope of $\beta$. The choice of a linear model with a zero intercept is justified by enforcing a physically meaningful condition of zero attenuation coefficient at zero frequency [49]. The units of $\alpha(f)$ and $\beta$ are Np/cm and Np/cm.MHz (or dB/cm.MHz), respectively. The factor 4 comes from signal's round-trip distance and amplitude squaring. In estimating the attenuation coefficient slope (ACS), to account for the diffraction term in equation (1), SDM uses a reference phantom to cancel out this term. To this end, ultrasound scans of a tissue-mimicking phantom with known acoustic properties is acquired using the same imaging system and setup as those used for soft tissue imaging. Under the assumption of homogeneity, the backscattering term is simplified as only a function of frequency, i.e. $B(f,z) \sim B(f)$. Using the subscript $s$ for the sample and $r$ for the reference phantom, we note $P_s(f) = P_r(f)$ and $D_s(f,z) \sim D_r(f,z)$. By employing the outlined models and assumptions above, the relative power spectral density of the sample to the reference phantom, $RS(f,z)$ (after taking the natural log of both sides) is derived from equation (2).

$$Ln\,[RS(f,z)] = -4(\alpha_s(f) - \alpha_r(f))(z - z_0) + Ln\left[\frac{B_s(f)A_{0,s}(f,z_0)}{B_r(f)A_{0,r}(f,z_0)}\right] \qquad (2)$$

Here, the second term on the right side is only a function of frequency and independent of depth $z$. The attenuation coefficient of the sample as a function of frequency, i.e. $\alpha_s(f)$, is estimated over a usable frequency range by the linear fit of equation (2) to the natural log of relative power spectra of the sample to phantom as a function of depth at each frequency component.

Backscatter intensity (power), $P$, is a quantitative measure of ultrasound image echogenicity that characterizes the strength of an echo coming from a tissue region. Depending on the size of scatterers, scatterer number density, impedance mismatch or the insonation angle, backscatter intensity can change. Omari and Varghese [59] showed that increasing the scatterer number density and size did not affect attenuation coefficient significantly when using the reference phantom method.

### B. Preclinical Porcine Model

Our longitudinal inflammation was conducted in a preclinical porcine model in a cohort of eight Sinclair mini pigs (four females and four males), as illustrated in Figure 2. It included six bi-weekly timepoints (termed as absolute timepoints), from week 0 (baseline, no inflammation) to week 10 (with inflammation). In each subject, three oral sites from each quadrant were included, resulting in a total of 12 oral sites per subject at a given timepoint. These three sites correspond to interdental gingival tissues at interproximal spaces between teeth (molars and/or premolars), which are listed below from posterior to anterior in the oral cavity:

Molar 1 -Distal (M1-Dis), Premolar 4 – Distal (PM4-Dis), Premolar 3 – Distal (PM3-Dis). This was a staggered study design in which a randomly selected oral quadrant from a pig was added to the study during a 10-week course (week 0 to week 10). The inoculation age, which is used to as the longitudinal inflammation timepoint, is defined for a quadrant as the difference between an absolute timepoint post-inoculation and the baseline timepoint of that quadrant. The distribution of added quadrants in each timepoint across the cohort are illustrated in Figure 2, where ellipses divided into four quadrants correspond to the animals' oral quadrants: left and right mandibular and maxillary. Pigs' quadrants were first scanned at their baselines, followed by the inoculation on the same visit. While imaging acquisitions was performed bi-weekly, the inoculation procedure was repeated weekly to maintain inflammation despite potential immune system recovery. To induce inflammation, two complementary approaches were applied simultaneously at each site, as presented in Figure 3 (a)-(c). The first approach aimed to promote plaque accumulation within the gingival sulcus by placing a silk ligature beneath the free gingival margin, a well-established procedure for inducing periodontal inflammation [60]. In the second approach, a bacterial agent known to induce periodontitis was injected into the gum through the space available at gingival sulcus, as shown in Figure 3 (c) to further accelerate the inflammation progression. Bacterial inoculation (Porphyromonas gingivalis, Fusobacterium nucleatum, and Treponema denticola) was performed at the rate of 5 $\mu L/min$ for one minute [61-63]. All scan acquisitions were conducted under general anesthesia while the animal was placed on an external heat source on the procedure table. Throughout the procedure, vitals including heart rate, respiration rate, and rectal temperature were continuously monitored. Before the study, all animals underwent teeth cleaning for enhancing consistency in the inflammation study design by minimizing dental plaque levels across animals prior to enrolments. The selection of porcine model here was driven by their similarities to human's oral histology and morphology [64]. The study was approved by Institutional Animal Care and Use Committee at the University of Michigan (PRO00010333). The University of Michigan's Animal Care & Use Program is accredited by AAALAC International.

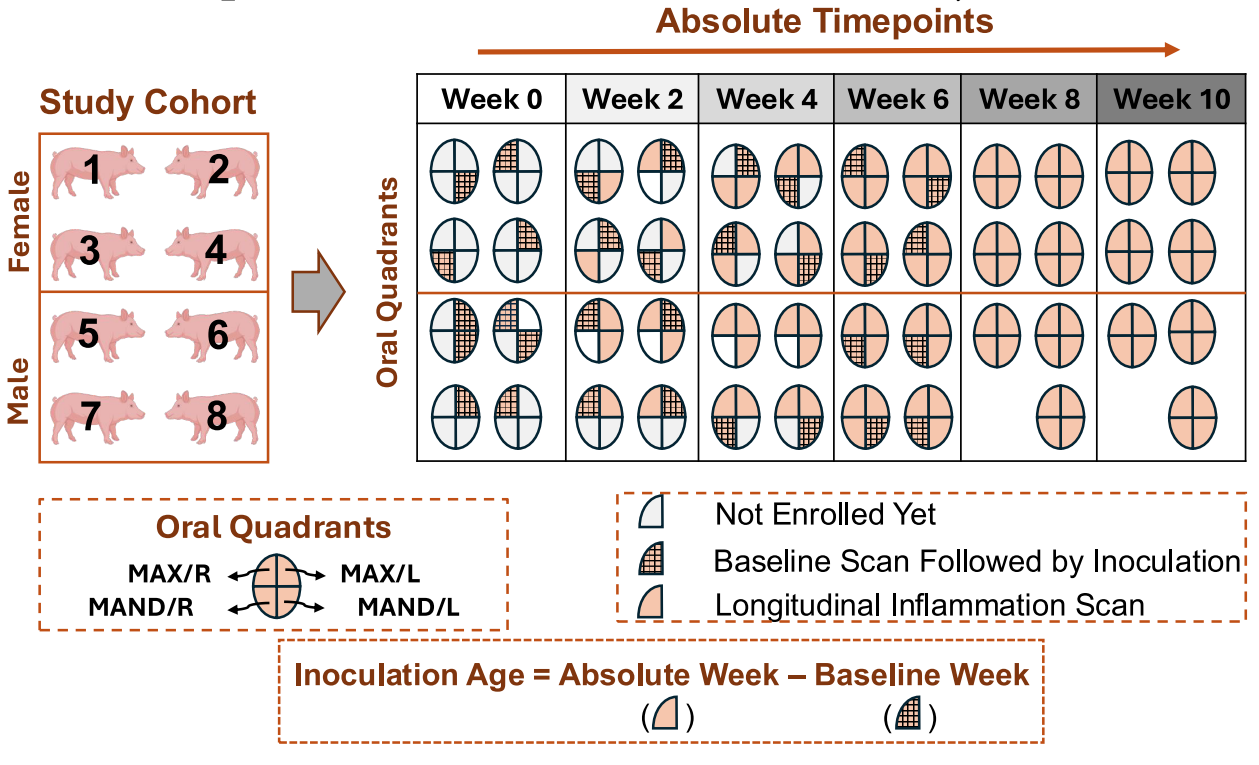

**Figure 2.** A schematic of the longitudinal staggered study design showing timepoints for inflammation experiments and imaging acquisitions based on quadrants' enrolment. Oral quadrants are represented as ellipses divided into four quadrants. Six absolute bi-weekly timepoints from week 0 to 10 are shown. Quadrant color pattern: white: not yet enrolled; orange with grid pattern: quadrant's baseline scan (followed by inoculation). Plain orange: the longitudinal inflammation scan post-baseline. The absence of subject 7 at week 10 denotes it was euthanized earlier in the experiment.

### C. Ultrasound Data Acquisition and Postprocessing

Ultrasound imaging was performed using an FDA-approved clinical ultrasound imaging system (ZS3, Mindray Innovation Center, NA, San Jose, CA, USA) paired with a small linear array transducer for intraoral scanning with the size of a toothbrush (L30-8, Mindray Innovation Center NA, San Jose, CA, USA), as shown in Figure 3 (d) – (f). The transducer has 128 elements with a center frequency of 18 MHz, operating at the transmit/receive frequency of 24 MHz for periodontal and phantom scans. The elevational focus of the transducer is located at the depth of 8 mm. A stand-off gel pad was mounted on the transducer surface to offset the periodontal tissues at the elevational focal region [30]. Ultrasound gel was also applied at the imaging oral site for acoustic coupling as well as minimizing any pressure applied by the operator onto the gum tissues during the imaging. Intraoral imaging of porcine periodontal tissues is presented in Figure 3 (g). For post-processing, the beam-formed IQ data were exported and RF (radiofrequency) data was obtained from IQ remodulation. B-mode image was reconstructed through envelop detection, log compression and scan conversion. A representative reconstructed B-mode image of periodontal tissues is shown in Figure 3 (h) with key oral anatomical structures annotated (epithelium (E), interdental gingiva (G), lining mucosa (M), alveolar bone (B), and crown and related clutter (C)). To estimate ACS, a rectangular region of interest (ROI) was manually placed within the gingival tissue region (interdental gingiva). ROIs were devoid of any regions with ultrasound imaging artifact such as reverberation and clutter. Epithelium and rete pegs were also excluded in ROI placement. BSIs were assessed within the same ROIs as those used for ACSs. In the spectral difference method, ACS within an ROI is obtained by segmenting RF data into multiple overlapping gates. The gate size was approximately eight times of the pulse length (64 $\mu m$) of the transducer, and the gate overlap ratio was 50% [57, 65]. Spectral representations of gated signals were obtained using the fast Fourier transform on which frequency smoothening were applied to reduce spectral noise. RF data were gates using a Hanning window to reduce spectral leakage. Additionally, consecutive scanlines were laterally averaged at each depth (with a three-scanline increment) to reduce the speckle noise effect within an ROI. A customized tissue-mimicking phantom was fabricated by an expert at the University of Wisconsin-Madison, Madison, WI, USA as the reference medium. This calibrated phantom had a speed of sound of 1540 m/s and ACS of ~0.7 dB/cm.MHz, and included beads with diameters ranging from 28-35 micrometer to provide scattering.

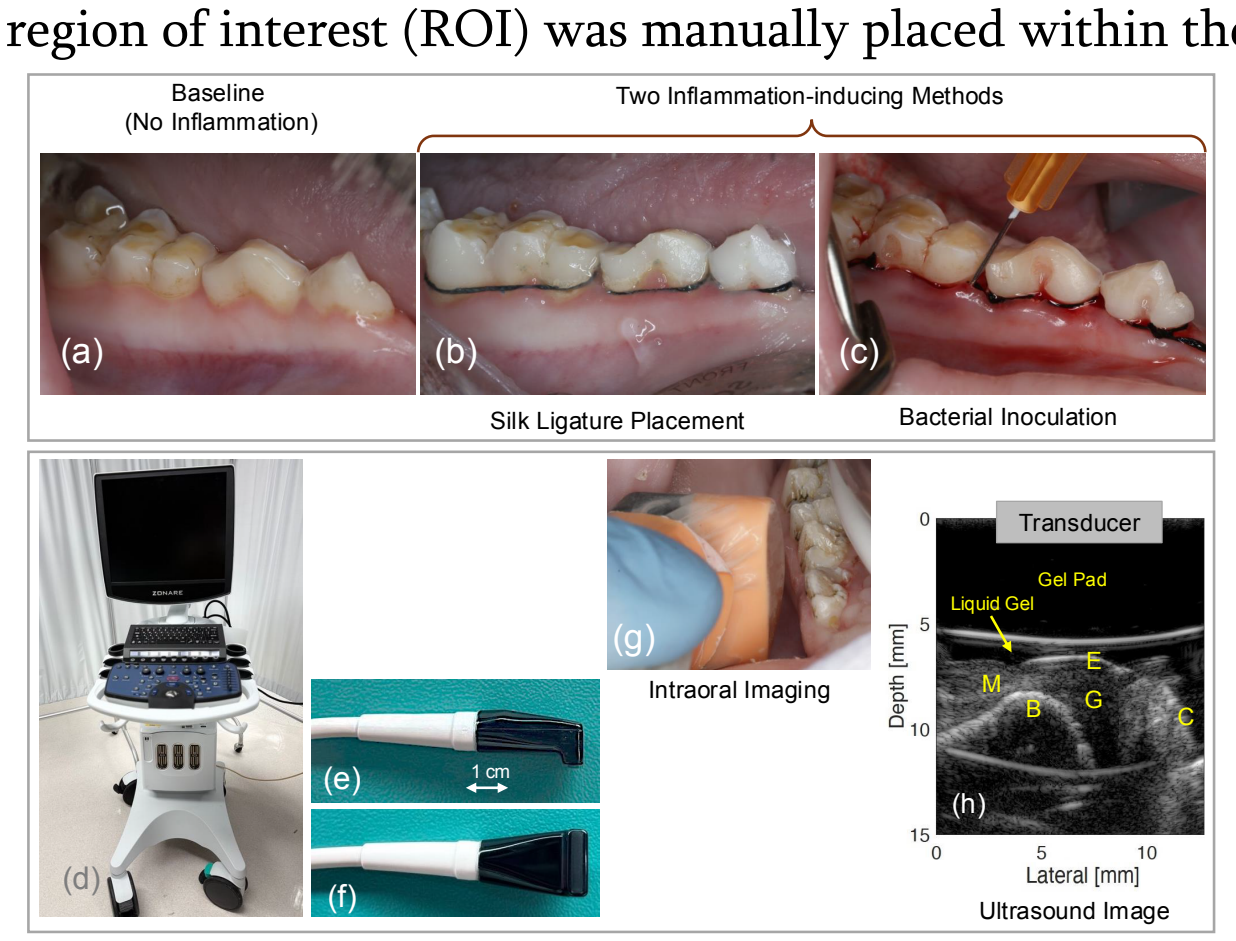

**Figure 3.** (a)-(c): Methods for inducing periodontal inflammation in a porcine subject. (a): baseline, (b) and (c): two inflammation induction approaches applied jointly. (b) the placement of silk ligatures sub-gingivally around teeth to increase plaque accumulation. (c) injection of the bacteria inoculation into the gum through gingival sulcus to induce periodontitis. (d) Ultrasound imaging system. (e) and (f) transducer's side and front views, respectively. (g) Intraoral imaging. (h) A representative B-mode image of pig periodontal tissues at PM4-Dis. Anatomical landmarks: E: Epithelium, G: Interdental Gingiva, B: Alveolar Bone, M: Mucosa, C: Crown and Related Clutter).

### *D. Statistical Analysis*

For statistical analysis, a linear mixed-effects model was implemented in MATLAB (equation (3)). Week was considered as the fixed effect while sex was considered as the covariate to adjust for possible male/female differences.

$$QUS \sim Week + Sex + Week * Sex + \ (1|Animal) \qquad (3)$$

The joint influence of week and sex was evaluated through the interaction term, i.e., $Week * Sex$ to determine whether longitudinal changes in ACS or BSI was different between males and females. Animal was included as a random effect to account for variabilities among subjects. Baseline was used as the reference timepoint. Statistical significance was defined as $p \leq 0.05$. If the sex effect and interaction term result in non-significant differences, the model is re-run by dropping the interaction term. We also tested the longitudinal changes of individual sexes, where the model simplifies into $QUS_{sex_i} \sim Week + \ (1|Animal)$.

## III. Results

### *A. Attenuation and Backscatter Intensity Analysis*

Figure 4 (a-e) show representative longitudinal intraoral B-mode images of a pig's oral tissues (Max-LM1-Dis) from week 0 (baseline) through four post-inoculation timepoints. Interdental gingival ROIs for QUS analysis are shown as white rectangular boxes. Figure 4 (f-j) presents corresponding attenuation coefficient plots versus frequency together with the linear one-parameter frequency model fitted to data, obtaining the ACS slope. The validation of high frequency ACS estimation techniques employed here was presented in our recent study through experiments on phantoms with known acoustic properties [49]. The summaries of longitudinal estimates of ACSs in three oral sites of M1-Dis, PM4-Dis, and PM3-Dis are shown in Figure 5 (a), (b), and (c), respectively. The effect of inflammation inoculation on tissue signature captured by ultrasound echo intensity are also presented as boxplots in in Figure 5 (d-f) for M1-Dis, PM4-Dis, and PM3-Dis, respectively. BSI was assessed within the same ROIs used for the ACS analysis. Here, pooled sexes are included in any of the timepoints. Using the mixed-effect model, three statistical significance tests was investigated.

**Sex Effect (Pooled Timepoints)**: When all timepoints were pooled into the analysis, no significant overall differences in ACS or BSI were observed between males and females at any of the three oral sites, with all $p > 0.05$.

**Sex-Specific Longitudinal Changes**: We assessed whether longitudinal changes in ACS or BSI differed by sex using the interaction term. No significant interaction was detected at any oral site (all $p > 0.05$) across all timepoints relative to their baselines, indicating that males and females exhibited similar longitudinal ACS or BSI changes at these timepoints. Since both sex-specific analyses above resulted in non-significant changes, the interaction term was dropped from the mixed effect model in pooled sex longitudinal analysis.

**Longitudinal Changes (Pooled Sexes)**: The statistical analysis of longitudinal ACS and BSI changes are reported for each oral site in Table 1. In M1-Dis and PM3-Dis, relative to the baseline, ACSs were significantly lower at all five inoculation timepoints ($p \leq 0.05$) when both sexes are pooled in the analysis. For PM4-Dis, we observe similar trend only at week 4. Looking at the median ACSs of these oral sites, either week4 or 6 were reported to show the smallest ACS estimates.In all oral sites, the BSIs of week 2 were significantly larger than. week 0. M1-Dis and PM3-Dis exhibited further longitudinal significance at week 4 with respect to their baseline, as well.

### *B. Sex as a Biological Variable in QUS Signatures*

Since sex may influence the inflammatory response of the subjects as well as underlying microstructures of oral tissues [66, 67], we employed sex as a variable to stratify the QUS estimates at longitudinal timepoints. Figure 6 presents ACS estimates stratified based on female (a-c) and male (d-f) cohorts at M1-Dis, PM4-Dis, and PM3-Dis in that order across all timepoints. Statistical significances between inflammation progression timepoints and baselines are reported in Table 1. In the female cohort, ACS at PM3-Dis across all inoculation timepoints were significantly lower than the baseline ($p \leq 0.05$) while in the male cohort, similar trend was observed at M1-Dis at all timepoints. M1-Dis of the female cohort exhibited ACS significance at week 2 and week 8 while PM3-Dis of male cohort showed such significance at week 6 and week 8. In both male and female cohorts, ACS of PM4-Dis did not show any statistical significance across inoculation timepoints.

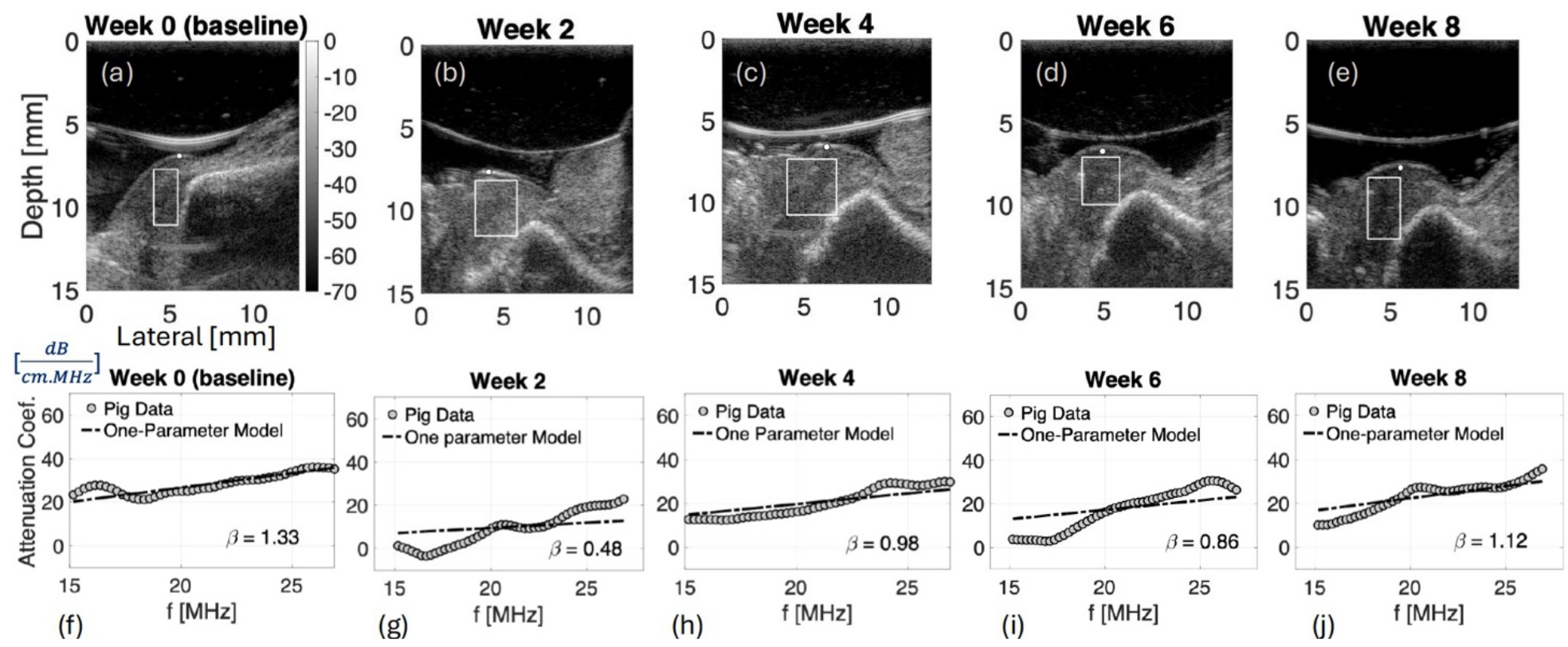

**Figure 4.** Representative of longitudinal interdental gingival scans and attenuation estimations within a porcine subject at five time points from pre-inoculation (week 0) to its four post-inoculation time points. (a-e): B-mode images with ROIs shown as white rectangular boxes. (f-j): corresponding attenuation coefficient plots and linear one-parameter model assessed based on the ROI's signal analysis.

**Table 1.** Pairwise comparison of the attenuation coefficient slope (ACS) and backscatter intensity (BSI) of interdental gingival tissues at each inoculation week with respect to the baseline for M1-Dis, PM4-Dis, and PM3-Dis. P-values for pooled sexes and sex-stratified QUS are reported. Asterisks: $p \leq 0.05$, ns: not significant.

| Oral Site | | M1 - Dis | | | | | PM4 - Dis | | | | | PM3 - Dis | | | | |
|---|---|---|---|---|---|---|---|---|---|---|---|---|---|---|---|---|
| **Inoculation Week** (w.r.t Week 0) | | 2 | 4 | 6 | 8 | 10 | 2 | 4 | 6 | 8 | 10 | 2 | 4 | 6 | 8 | 10 |
| ACS (Pooled Sexes) | | 0.000 * | 0.000 * | 0.000 * | 0.000 * | 0.006 * | 0.757 ns | 0.050 * | 0.095 ns | 0.258 ns | 0.888 ns | 0.001 * | 0.003 * | 0.000 * | 0.000 * | 0.003 * |
| BSI (Pooled Sexes) | | 0.000 * | 0.003 * | 0.268 ns | 0.417 ns | 0.609 ns | 0.000 * | 0.152 ns | 0.245 ns | 0.151 ns | 0.763 ns | 0.000 * | 0.000 * | 0.154 ns | 0.363 ns | 0.389 ns |
| ACS | Females | 0.029 * | 0.065 ns | 0.062 ns | 0.013 * | 0.169 ns | 0.889 ns | 0.249 ns | 0.640 ns | 0.924 ns | 0.335 ns | 0.006 * | 0.005 * | 0.000 * | 0.000 * | 0.013 * |
| | Males | 0.000 * | 0.000 * | 0.000 * | 0.002 * | 0.005 * | 0.750 ns | 0.116 ns | 0.061 ns | 0.128 ns | 0.457 ns | 0.066 ns | 0.132 ns | 0.011 * | 0.009 * | 0.068 ns |
| BSI | Females | 0.002 * | 0.021 * | 0.362 ns | 0.462 ns | 0.675 ns | 0.000 * | 0.022 * | 0.072 ns | 0.003 * | 0.488 ns | 0.000 * | 0.002 * | 0.065 ns | 0.286 ns | 0.513 ns |
| | Males | 0.000 * | 0.013 * | 0.322 ns | 0.533 ns | 0.547 ns | 0.004 * | 0.592 ns | 0.648 ns | 0.913 ns | 0.991 ns | 0.144 ns | 0.031 * | 0.693 ns | 0.709 ns | 0.498 ns |

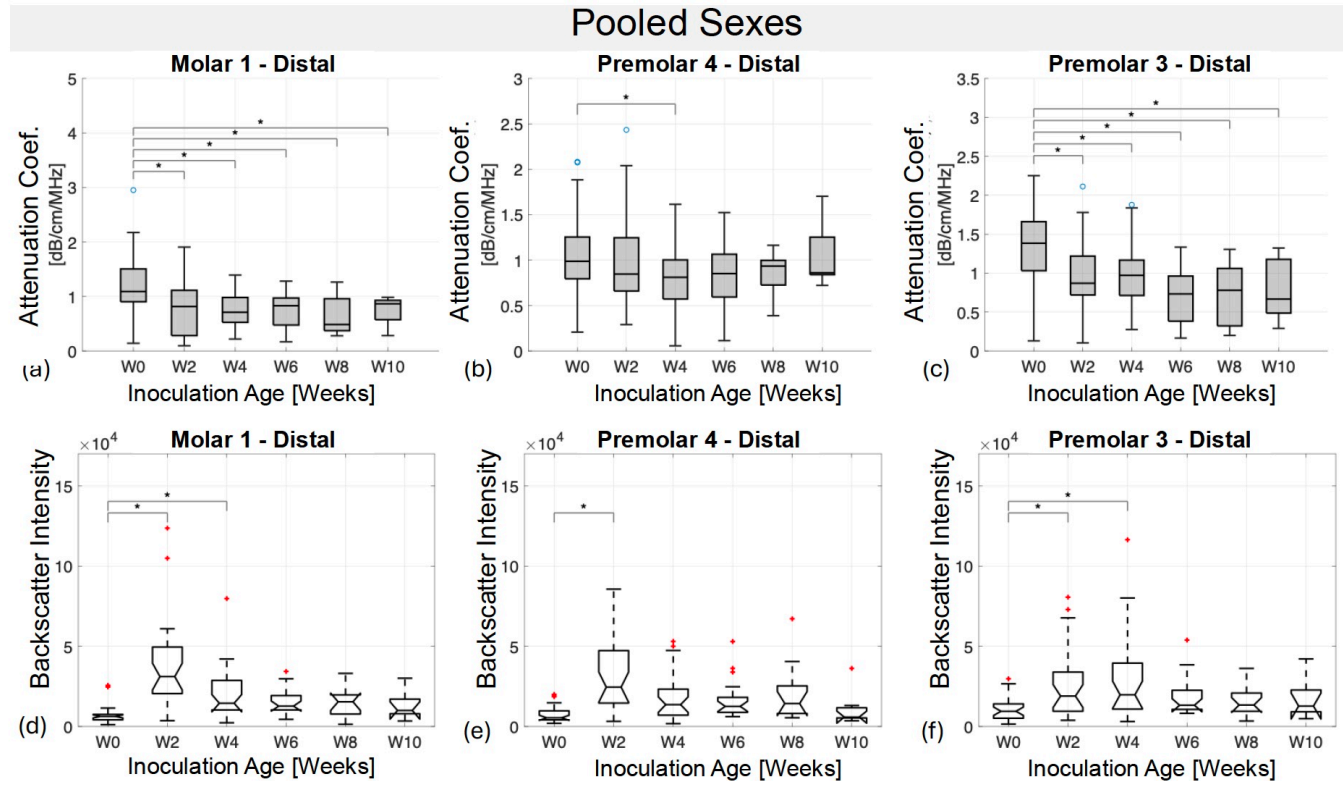

**Figure 5.** Longitudinal comparison of ASC and BSI (pooled sexes). (a) and (d): M1-Dis; (b) and (e): PM4-Dis, and (c) and (f): PM3-Dis. Asterisk denotes a statistical significance ($p \leq 0.05$) (Blue/red symbols: outliers).

Sex-based stratifications of BSI values across inoculation timepoints are shown in Figure 7 (a)-(c) for females and Figure 7 (d)-(f) for males with p-values reported in Table 1. In the female cohort, relative to baseline, BSIs were significantly higher at weeks 2 and 4 across M1-Dis, PM4-Dis, and PM3-Dis. In this cohort, further statistical significance was observed at week 8 (PM4-Dis), as well. In the male cohort, the molar site (M1-Dis) was consecutively distinct at week 2 and 4 relative to baseline. Looking at both premolars of this cohort, either of these timepoints appeared as distinct: PM4-Dis at week 2 and PM3-Dis at week 4.

### *C. QUS-Based 2D Inflammation Classification*

Given the complex nature of pathological conditions such as inflammation, a multiparametric approach may offer enhanced characterization capabilities. Thus, we further look at the separations of inflammation timepoints through the 2D QUS parameter space of ACS versus BSI. Considering week 2 (which demonstrated more frequent significance from corresponding baselines across all oral sites), this timepoint was selected for the 2D classification, as shown in Figure 8 (a-c) for M1-Dis, PM4-Dis, and PM3-Dis respectively (pooled cohort analysis). Here, a linear boundary was also optimized based on maximum separation of baseline from the week 2. The 2D classification performances were 92% for M1-Dis, 82% for PM4-Dis, and 74.2% for PM3-Dis. Some outliers are not visible in 2D plots to provide a zoomed-in view of the classification results.

## IV. Discussion

To our knowledge, this is the first study that investigates the effect of periodontal inflammation on estimated QUS parameters of ACS and BSI. In an in vivo longitudinal study at baseline and five inflammation timepoints, we presented a comprehensive QUS analysis of three oral sites around molars and premolars: PM3-Dis, PM4-Dis, and M1-Dis. Sex was considered as a covariate (joint effect of sex and timepoint as well as sex-based stratification) to account for possible differences between males and females in terms of oral tissue microstructure, host's immune response and tissue remodelling with inflammation inoculation. Different teeth (molars vs. premolars) were also evaluated since structural differences between teeth may influence QUS parameters. Our findings showed that at weeks 2 and 4 of inflammation inoculation, both ACS and BSI were consistently and significantly different than the baseline at all oral sites for the pooled sexes ($p \leq 0.05$), except for ACS at week 2 and BSI at week 4 for PM4-Dis. PM4-Dis exhibited less distinct changes than the other two oral sites. In sex-based analysis, PM4-Dis appeared overall non-responsive in association of ACS with inflammation inoculation. We highlight that PM4-Dis is located adjacent to both molar and premolar teeth structures whereas PM3-Dis and M1-Dis were each bordered exclusively by either premolars or molars, respectively. Females showed slightly more response to inflammation than the males in BSI and ACS assessment. Inflammation inoculation appeared to decrease ACS values while increasing the measured BSI within investigated ROIs.

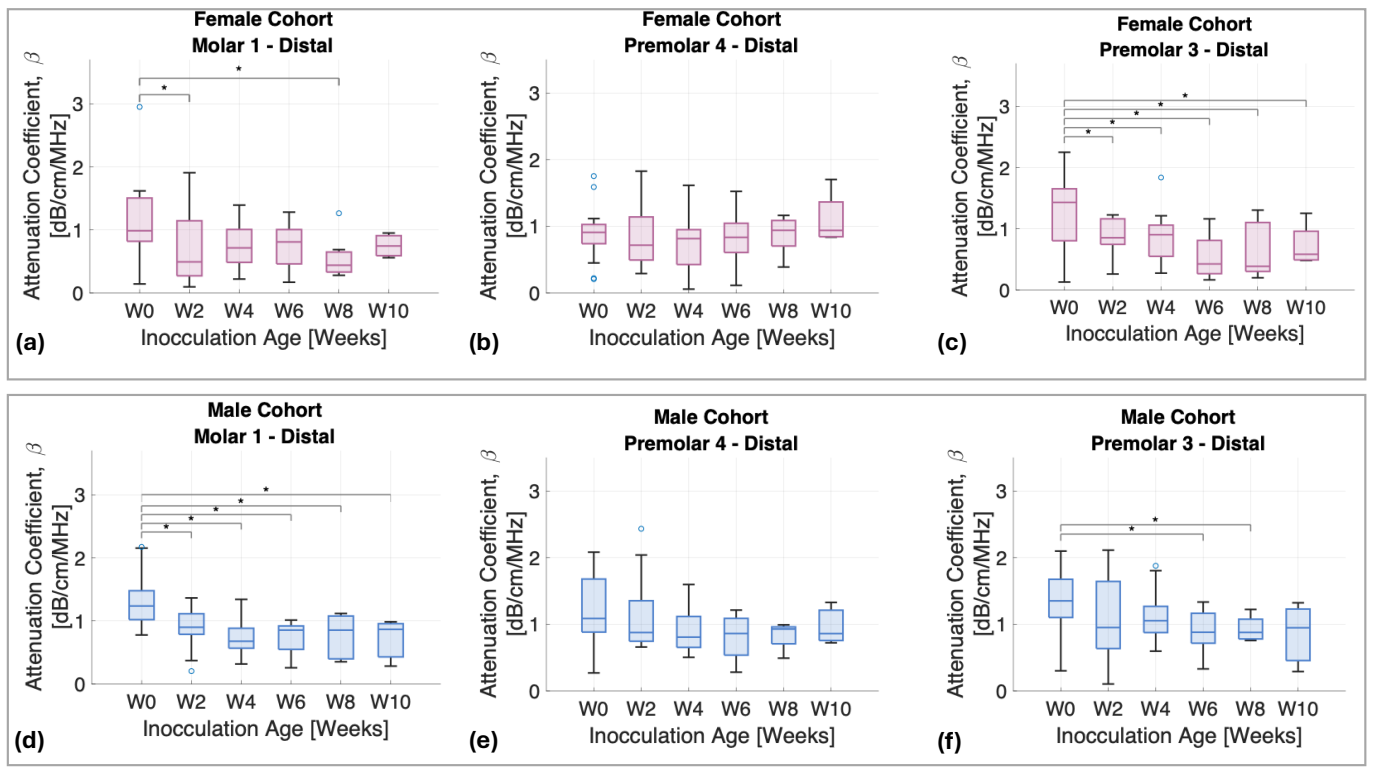

**Figure 6.** Longitudinal attenuation coefficient stratified based on the sex at three interproximal oral sites. Top row (a-c): male cohort, Bottom row (d-f): female cohort. Blue circles show outliers. Asterisks denote statistical significances.

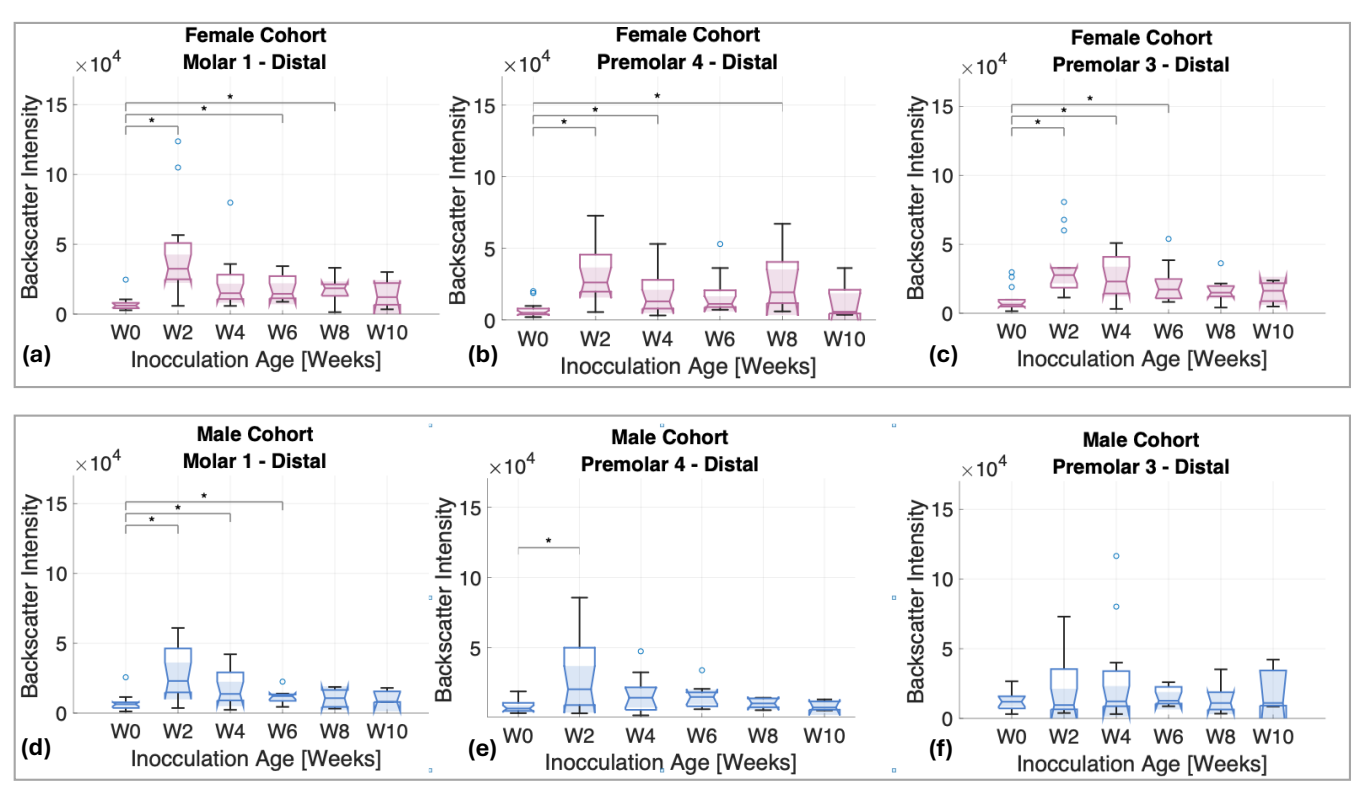

**Figure 7**. Longitudinal backscatter intensity stratified based on sex at three interproximal oral sites. (a-c): female cohort, (d-f): male cohort. Asterisks denote statistical significances. Blue symbols show outliers.

An explanation for the observed ACS decrease with inflammation inoculation could be a possible increase in edema (i.e. elevated water content and tissue swelling) associated with inflammation. Since ultrasound waves attenuates less in water than tissue, even a small increase in fluid content, i.e. the inflammation-induced edema, maycontribute to lowering the attenuation estimates (ACS of water is around 0.001 dB/cm.MHz [68]) compared with the higher ACS of healthy oral soft tissues being around ~1 dB/cm.MHz [49]). A lower attenuation indicates lower overall signal decay with ROI depth across inoculation timepoints compared with the baseline. Consequently, this may explain the lower mean echo intensity at baseline compared to higher echo intensity at inoculation timepoints (Figure 5), particularly since BSI was measured within the same ROIs used for ACS estimation.

To provide further insights into our longitudinal QUS measurements with inflammation, we discuss our findings with a context of a recent study from our group on the same porcine study design as in this study but different ultrasonography analysis. Samal et. al [39] studied changes in gingival blood flow velocity and power with inflammation at mid-facial (Mid) sites (p-values were reported for three of the timepoints: weeks 2, 4 and 6). They showed a significantly higher velocity estimates for three inflammation timepoints at M1-Mid and PM3-Mid in both cohorts and at PM4-Mid in female cohort. Only week 2 of male cohort at PM4-Mid were significant. Our findings comparing ACS or BSI variations across inoculation week showed an overall alignment with the latter study. In terms of longitudinal changes, our ACS estimates also demonstrated separations of weeks 2, 4, 6, as well as weeks 8, and 10 relatives to baseline at PM3-Dis and M1-Dis except for PM4-Dis being significant only at week 4 in our pooled sex analysis. Our BSI estimate also showed such significance at week 2 and 4 except for PM4-Dis at week 4. Our findings on sex-based stratifications of BSI being consistently significant at weeks 2, 4 and/or 6 versus baseline for females at all oral sites were mostly aligned with the distinction of their color flow velocity. Moreover, our decreased ACS trend and elevated mean BSI may be interconnected to their increased blood flow velocity with inflammation. We note that an increased blood flow in response to inflammation may result in vasodilation, leading to vascular permeability (where the blood plasma leaks out and forms edema), reducing ACS and BSI, as explained before.

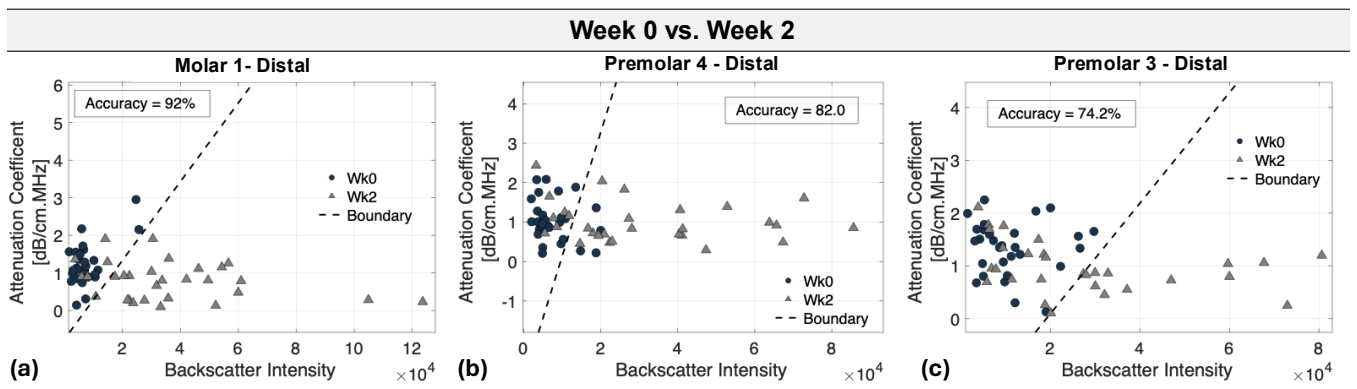

**Figure 8.** Classification of the inoculation timepoint of week 2 from baseline using QUS 2D parameter space. (a) M1-Dis, (b) PM4-Dis, and (c) PM3-Dis.

Our findings suggest that both sex and oral sites may be among factors confounding the inflammation response in oral soft tissues. The 2D classification accuracy for separating week 0 from week 2 (appearing distinct more often) showed that both QUS parameters contributed to the separation, highlighting the advantage of employing a multiparametric approach in improved oral inflammation characterization. As a note on the relatively large variations in QUS estimates, it may be attributed to the difference in ROI sizes across scans driven by the available scanned tissue areas, as well as the heterogenous nature of gingival tissues with numerous fiber arrangements (See the first figure in [47]). We highlight that the backscatter signal coming from tissues depend on the relative density and compressibility of tissue microstructures in comparison to their background, orientation of such fibers with respect to the imaging transmit signal as well as tissue vascularization.

An ACS estimation based on analysis of pixel intensities of a compressed B-mode image without accounting for system effects [46] deviates from standardized approaches and does not yield accurate ACS values. The ACS in this study was assessed using standardized techniques in physical acoustic through implementing spectral difference approach on uncompressed echo signal [49].

The BOP approach to assess gingival inflammation could not be adopted as a clinical inflammation indicator in this study. This was limited by one of the inflammation-inducing approaches where ligatures were pushing into the gingival sulcus to promote local bacteria accumulation and inflammation. However, the visual assessment of the oral inflammation was assessed by periodontists in weekly bacterial inoculations as an overall examination approach. While BOP has many limitations including subjectivity and false positive outcomes future investigations may also focus on BOP implementation in the study design. Moreover, other QUS-based techniques to study inflammation, such as speckle modelling are yet to be explored. Another promising research area is the multiparametric QUS inflammation monitoring. Future studies should focus on expansion of QUS analysis to clinical studies and investigate variations of ACS and BSI in response to periodontitis and gingivitis compared with healthy gums. Due to the staggered nature of this longitudinal study, lower number of ultrasound scan acquisitions existed in final inoculation timepoints, particularly when stratified by sex, which is a limitation of this study. Future work may also focus on higher enrolments.

## V. Conclusion

This study suggests the promising potential of intraoral ultrasonography paired with QUS-based techniques for the assessment and quantification of oral inflammation. To our knowledge, this is among early investigations into oral inflammation monitoring using QUS-based analysis. We characterized ACS and BSI of interdental gingival tissues with inflammation inoculation timepoints compared with the healthy tissues. We also investigated stratification of QUS parameters' variations with inflammation based on sex and oral sites (three interproximal oral sites around molars and premolars). Overall, the ACS decreased with inflammation inoculation within the investigated ROIs while the BSI measured from the same ROIs appeared to increase at inflammation timepoints. We also presented the potential of multiparametric QUS-based analysis for inflammation assessment. Ultrasonography may also offer promises in assessment of other pathological conditions.

## Disclosure

The authors declare no conflicts of interest associated with this work.

## Acknowledgement

This work was supported by the National Institutes of Health, award numbers R21DE029005 (O.D.K and H-L.C.) and F32DE034986 (D.P.). The authors acknowledge Dr. Kerby Shedden for the valuable statistical consultation and insightful guidance.